\documentclass[%
reprint,
superscriptaddress,
amsmath,amssymb,
aps,
pra,
floatfix,
]{revtex4-2}
\usepackage[colorlinks=true]{hyperref}
\usepackage{graphicx, xcolor}
\usepackage{braket}
\usepackage{subcaption}
\usepackage{array}
\usepackage{makecell}
\newcolumntype{P}[1]{>{\centering\arraybackslash}p{#1}}
\newcolumntype{M}[1]{>{\centering\arraybackslash}m{#1}}

\renewcommand{\eqref}[1]{Eq.~(\ref{#1})}

\newcommand{\ip}{\mathrm{I}_\mathrm{p}}
\newcommand{\up}{\mathrm{U}_\mathrm{p}}
\newcommand{\pb}{\mathbf{p}}

\newcommand{\rb}{\mathbf{r}}

\newcommand{\Ab}{\mathbf{A}}

\begin{document}

\title{Time Correlation Filtering Reveals Two-Path Electron Quantum Interference in Strong-Field Ionization}
\author{Nicholas Werby}
\email{nwerby@stanford.edu}
\affiliation{Stanford PULSE Institute, SLAC National Accelerator Laboratory\\
2575 Sand Hill Road, Menlo Park, CA 94025, USA}
\affiliation{Department of Physics, Stanford University, Stanford, CA 94305, USA}

\author{Andrew S. Maxwell}
\affiliation{Department of Physics and Astronomy, Aarhus University, DK-8000 Aarhus C, Denmark}
\affiliation{Institut de Ciencies Fotoniques, The Barcelona Institute of Science and Technology, 08860 Castelldefels (Barcelona), Spain}
\affiliation{Department of Physics \& Astronomy, University College London, Gower Street,  London,  WC1E 6BT, United Kingdom}

\author{Ruaridh Forbes}
\affiliation{Stanford PULSE Institute, SLAC National Accelerator Laboratory\\
2575 Sand Hill Road, Menlo Park, CA 94025, USA}
\affiliation{Department of Physics, Stanford University, Stanford, CA 94305, USA}
\affiliation{Linac Coherent Light Source, SLAC National Accelerator Laboratory, Menlo Park, California 94025, USA}

\author{Carla Figueira de Morisson Faria}
\affiliation{Department of Physics \& Astronomy, University College London, Gower Street,  London,  WC1E 6BT, United Kingdom}

\author{Philip H. Bucksbaum}
\affiliation{Stanford PULSE Institute, SLAC National Accelerator Laboratory\\
2575 Sand Hill Road, Menlo Park, CA 94025, USA}
\affiliation{Department of Physics, Stanford University, Stanford, CA 94305, USA}
\affiliation{Department of Applied Physics, Stanford University, Stanford, CA 94305, USA}


\date{\today}

\begin{abstract}

Attosecond dynamics in strong-field tunnel ionization are encoded in intricate holographic patterns in the photoelectron momentum distributions (PMDs). These patterns show the interference between two or more superposed quantum electron trajectories, which are defined by their ionization times and subsequent evolution in the laser field. We determine the ionization time separation between interfering pairs of electron orbits by performing a differential Fourier analysis on the measured momentum spectrum. We identify electron holograms formed by trajectory pairs whose ionization times are separated by less than a single quarter cycle, between a quarter cycle and half cycle, between a half cycle and three fourths of a cycle, and a full cycle apart. We compare our experimental results to the predictions of the Coulomb quantum orbit strong-field approximation (CQSFA), with significant success. We also time-filter the CQSFA trajectory calculations to demonstrate the validity of the technique on spectra with known time correlations. As a general analysis technique, the filter can be applied to all energy- and angularly-resolved datasets to recover time correlations between interfering electron pathways, providing an important tool to analyze any strong-field ionization spectra.

\end{abstract}


\maketitle
\section{Introduction}
\label{sec:Intro}

Laser-induced strong-field ionization (SFI) is a powerful source of attosecond electron dynamics. Field-ionized electrons can be driven by the laser into many different types of orbits about the parent ion before escaping and being captured by a detector.
Angle-resolved SFI photoelectron momentum distributions such as those captured by velocity map imaging (VMI) display complex patterns that have prompted significant study \cite{okino_direct_2015,xu_self-imaging_2010,willenberg_holographic_2019,tan_time-resolving_2019,he_direct_2018,walt_dynamics_2017,zhou_near-forward_2016,meckel_signatures_2014,marchenko_criteria_2011}. Many of these patterns are thought to be due to quantum interference between pairs of electrons that take different pathways but reach the same spot on the imaging detector. These trajectories depend on the optical field phase at the moment of ionization, as well as the temporal shape of the ionizing field. The study of these interfering trajectories, and the momentum patterns they produce, is called electron holography, and has gained significant attention in the past decade \cite{huismans_time-resolved_2011,hickstein_direct_2012,meckel_signatures_2014,faria_it_2020}. 

As a field, electron holography is predominantly theory-based rather than experimental. Trajectory-based calculations are relatively easy to manipulate to study relevant pathways that give rise to specific interference features. 
In contrast, holographic analysis of experiments is more challenging because the laser field phase at the moment of electron ionization is generally not an experimental observable, nor is it possible to eliminate specific trajectories in an experiment. Recently, some notable progress has been made using the ``phase of the phase'' technique, in which the phase-shifted second harmonic of the fundamental laser is used to shape the ionizing field to emphasize field ionization at specific ionization phases \cite{wurzler_accurate_2020, tan_resolving_2021, skruszewicz_two-color_2015}. This has promise for extracting ionization times for some parts of the spectra such as classical cusps and regions of back-scattering \cite{skruszewicz_two-color_2015, wurzler_accurate_2020}, but cannot reveal features composed of multiple contributing trajectories with different ionization times \cite{tan_resolving_2021}, which create the features of greatest interest in photoelectron holography. 

Fourier analysis may be performed on any energy-resolved coherent power spectrum to reveal temporal \textit{autocorrelations} of coherent interfering pathways independent of theory \cite{10.1007/BF02546511}. Here, we show that filtered time correlation analysis can be applied to angle-resolved SFI electron momentum distributions (PMDs) to reveal the time differences between the launch times of interfering trajectory pairs, and thus may provide quantitative empirical validation for the primary predictions of electron holography theory. By extracting the features produced by specific trajectory time separations, we can directly classify trajectory pairs composed of two rescattered orbits, two direct orbits, or one of each. This has significant application for understanding strong-field interactions, since rescattered electron dynamics encode information about transient structure in the ionized target \cite{chen_analysis_2007, yuan_exploring_2021, ito_rescattering_2018, chen_role_2015, xu_self-imaging_2010, morishita_retrieval_2009}. Identifying and isolating features produced by these types of trajectories can, therefore, be exploited for analyzing molecular movies in energy- and angularly-resolved pump-probe experiments \cite{PhysRevLett.125.123001}. 


The remainder of this paper is laid out in the following way. 
Section \ref{Sec:Exp_Methods} briefly presents the experimental set-up for momentum imaging and the basic data processing required for the time correlation filter. Section \ref{Sec:Theory_Methods} introduces a computational model for strong-field ionization. In Section \ref{Sec:Discussion} we show the model predictions, which are in significant agreement to the analyzed time correlations. Atomic units are used throughout, except where otherwise indicated.

\section{Experimental Methods and Analysis}
\label{Sec:Exp_Methods}

\subsection{Description of Experiment}
\label{Sec:Exp_Description}

\begin{figure}
	\centering
	\includegraphics[width=1\linewidth]{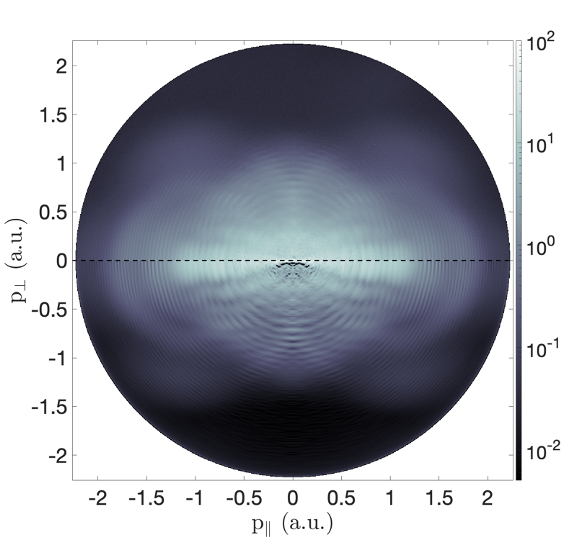}
	\caption{Top: The raw PMD of the argon gas dataset used throughout this paper. Bottom: The same spectrum after Abel inversion (see text).}
	\label{fig:Exp_Spectrum}
\end{figure}

Holography patterns are expected to occur in the SFI electron momentum distributions for all gases. Here we show a well-studied example, the PMD of argon gas photoionized by 800~nm infrared laser pulses generated by a 1~kHz Ti:sapphire laser (Fig. \ref{fig:Exp_Spectrum}). This spectrum was obtained using 40~fs full width at half maximum (FWHM) linearly polarized pulses with peak intensity of 225$\pm12.5$~TW/cm$^2$. The intensity was determined by the ponderomotive cutoff visible in the PMD spectrum, as well as by fits to the calculated location of nodes along the spider-leg shaped interference features \cite{werby_dissecting_2021}. A pulsed and skimmed beam of argon gas intersected the interaction region in an ultrahigh vacuum. Electrons were detected in a standard velocity-map imaging (VMI) apparatus and recorded on an intensified phosphor viewed by a camera. For each laser shot, on-the-fly peakfinding was employed to record the pixel locations of each electron impact. These were summed to form the final spectrum.

2.4 billion electron counts were recorded in the spectrum presented in Fig.~\ref{fig:Exp_Spectrum}. The top half shows the raw summed electron counts in the VMI, projecting the three-dimensional Newton sphere of final electron momenta onto the detector. Since SFI with linearly polarized light has cylindrical symmetry about the laser polarization ($\parallel$) axis, the cross section of the Newton sphere containing the polarization axis and any perpendicular axis displays all the relevant dynamics. To extract the momentum dependence, we apply polar onion peeling inversion \cite{roberts_toward_2009}. Beginning at the outermost radius on the raw image, polar onion peeling fits a Legendre decomposition to the photoelectron angular distribution (PAD) at that radius,  
\begin{equation}
    I(\theta, p_r) = C_0(p_r)\sum_{n_{\mathrm{even}}} \beta_n(p_r)P_n(\cos(\theta)),
    \label{eqn:L_Decomposition}
\end{equation}
where $I(\theta, p_r)$ is the PAD fit for polar angle $\theta$ and radius $p_r$, $P_n(\cos(\theta))$ are the Legendre polynomials of order $n$, $C_n(p_r)$ is the nth order Legendre coefficient of the fit, and $\beta_n(p_r)$ are the anisotropy parameters defined as $\beta_n(p_r) = \frac{C_n(p_r)}{C_0(p_r)}$. We fit up to 42nd Legendre order, which is the order at which the variance between the residuals of additional orders is minimized. By convolving this fit about the polarization axis, projecting back into the detector plane, and subtracting the projection from the image we effectively ``peel'' the contribution of that PAD from the remainder of the spectrum. Iterating for each radius results in the fully inverted spectrum, shown in the bottom half of Fig. \ref{fig:Exp_Spectrum}.

\begin{figure*}
	\centering
	\includegraphics[width=0.85\linewidth]{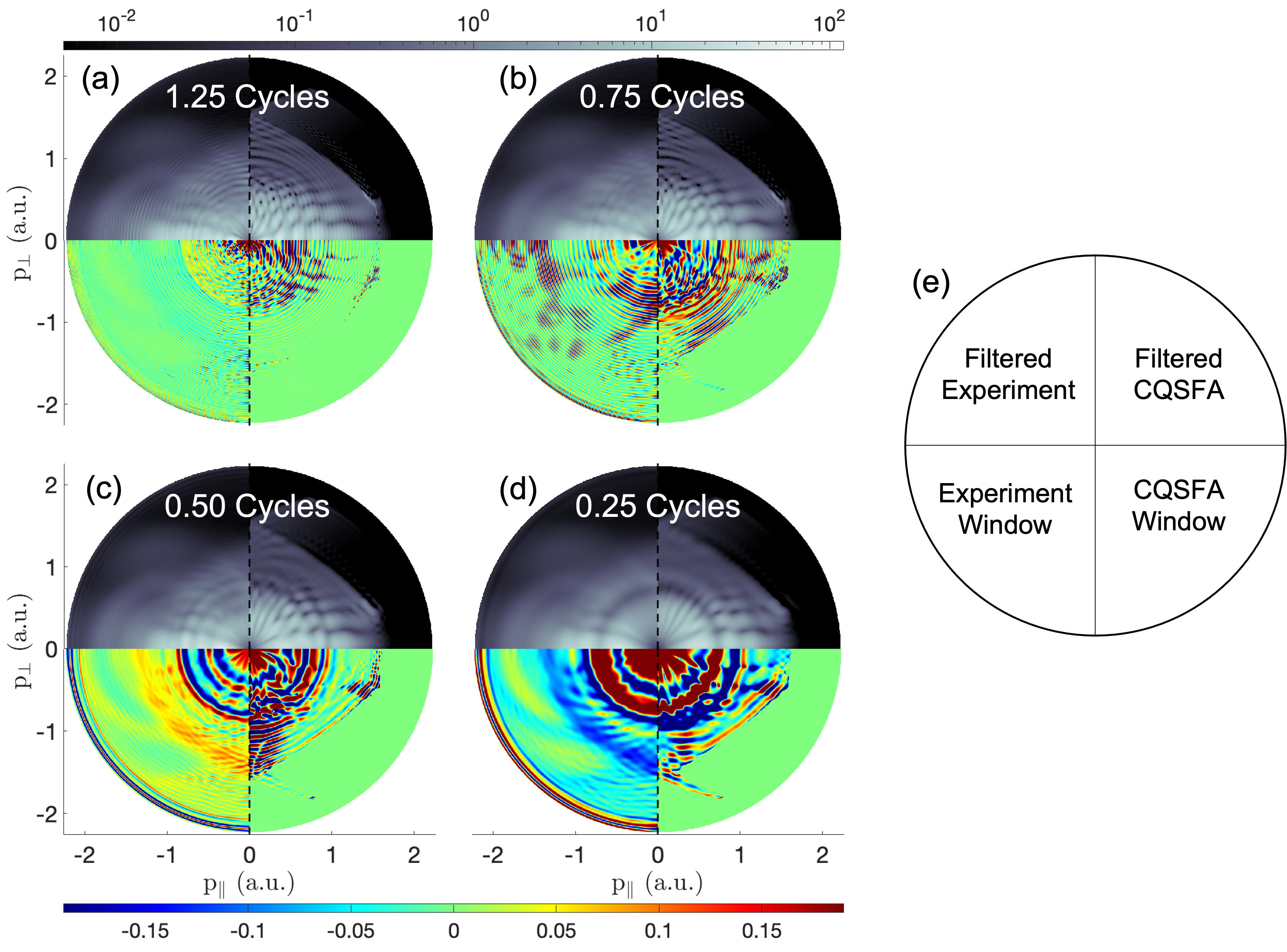}
	\caption{The product of the time correlation filter applied at successive filter rolloffs indicated by the annotation at top center. The left half of each panel displays the experimental data, whereas the right half shows the full CQSFA calculation with all orbits (see Section \ref{Sec:Theory_Methods}). The top half of each panel is the filtered spectrum, and the bottom halves show the normalized residuals (filter windows, see text) between the previous step of filtering and the step indicated by the annotation on the upper panels. Panel (e) is a reference schematic for the previous panels. The colorbar for the top half of each panel is at the top, while the colorbar for the bottom half of each panel is at the bottom. 
	}
	\label{fig:Exp_Succession}
\end{figure*}

\subsection{The Time Correlation Filter}
\label{Sec:TC_Filter}


The cylindrically-symmetric momentum spectrum $I(\theta, p_r)$ can be re-expressed as an angle-resolved electron power spectrum by converting the radial momentum dependence to kinetic energy for each anisotropy parameter generated in polar onion peeling. The cosine transformation of these parameters then yields a temporal autocorrelation of the electron trajectory currents that produce the spectrum \cite{werby_disentangling_2021}: 
\begin{equation}
    \frac{1}{2\pi}\int_{0}^{\infty}\beta_n(E)e^{iEt}dE = \int y_n(\tau-t)y_n(\tau)d\tau.
\end{equation}
Here $\beta_n(E)$ are the anisotropy parameters defined at the electron energy $E$, $y_n(\tau)$ is the complex electron yield produced by all trajectories ionized at time $\tau$, and $t$ is the time \textit{difference} between pairs of orbits. The integral on the right can be interpreted as the complete set of all pairs of trajectories with defined time difference $t$. Thus, the cosine transform of the anisotropy parameters directly reflect the time separation between interfering orbits, with its amplitude reflecting the density of electron trajectories interfering with that separation. 

In semi-classical trajectory-based interpretations of SFI, interference fringes will form with close to constant separation in energy based on the time difference between the launch time of the interfering trajectories (Note: see subsection \ref{sec:Limitations}) \cite{maxwell_carpet_2020, faria_it_2020}. Recognizing this, we apply a low-pass filter to the transformed $\beta_n(E)$ and then transform back into momentum space to eliminate the contributions from all interfering orbits separated by more than the filter rolloff. Filtering the data at a series of rolloffs grants us a three dimensional dataset, with the time separation as our third parameter. This has significant impact on the kinds of features we are able to resolve in SFI spectra. 

For the data presented here, we use a Kaiser window, finite impulse response low-pass filter. The Kaiser window is used to ensure the response is maximally flat in the passband. The attenuation in the stopband is set to 20 decibels, and the transition width is set to 0.1 cycles. Due to the numerical resolution of our dataset, in the filtered data we present we are limited to a lowest rolloff of 0.25 cycles. A movie of the filtered data at 0.01 cycle steps is available in the supplementary materials. 

In previous works, we primarily used the time correlation filter to filter at one laser cycle of time separation. This effectively unlocked the sub-cycle electron momentum distribution by eliminating the pervasive ring-like interference structure caused by above-threshold ionization (ATI) \cite{werby_disentangling_2021}. Simply removing this interference pattern allowed for significant quantitative comparisons with calculation that had previously been completely obscured. However, the time correlation filter can also be used to isolate and examine specific holographic interference patterns, and experimentally verify the predictions of calculations as to the class of trajectories that form them.

By observing the residuals of spectra with filter rolloffs lower than one cycle, we can directly observe the time separation leading to various holographic patterns. This is very significant. Knowing the time separation allows us to demonstrate experimentally whether interfering trajectories are launched within the same quarter cycle, on adjacent rising and falling edges of the laser field, on opposing half cycles, or separated by a full cycle. These characterizations are extremely useful for identifying the class of trajectories being observed. 

In Fig. \ref{fig:Exp_Succession} we present the results of time correlation filtering at different filter rolloffs. Each panel, indicated by (a)-(d), displays four spectra for comparison, one in each quadrant. It should be noted that these PMDs are symmetric in each quadrant, so there is no difference in the negative or positive components of each momentum axis. Each panel is divided in the same way, with experimental data presented on the left half compared to a calculation using the Coulomb quantum orbit strong-field approximation (CQSFA) on the right. The calculations will receive further context in the following section. The top halves of each panel show the filtered PMDs at the indicated rolloff in units of laser field cycles. Looking at these filtered PMDs alone, it can be difficult to isolate the holographic features that have been removed at each step. To specifically extract these features we employ the normalized residual defined as

\begin{equation}
    \delta Y_{i,j}(p_\parallel, p_\perp) = \frac{Y_{i}(p_\parallel, p_\perp)- Y_{j}(p_\parallel, p_\perp)}{Y_{i}(p_\parallel, p_\perp)+Y_{j}(p_\parallel, p_\perp)},
\end{equation}
where $Y(p_\parallel, p_\perp)$ is the PMD, and $i$ and $j$ indicate the two PMDs being compared. The bottom halves of each panel show the normalized residuals between the previous less filtered step and the indicated step. In \ref{fig:Exp_Succession}(a), the normalized residual is taken with respect to the unfiltered PMD. The structures in each residual panel are extracted quantum interference patterns whose trajectory pair have launch-time separation within the residual window. As a clarifying example, the normalized residual in \ref{fig:Exp_Succession}(c) uses the PMD in \ref{fig:Exp_Succession}(b) as $Y_i$ and the PMD in \ref{fig:Exp_Succession}(c) as $Y_j$, and so displays an ionization window of 0.5 cycles to 0.75 cycles. In general, we call these residuals ``filter windows.''


\section{Theoretical Methods}
\label{Sec:Theory_Methods}

We employ the CQSFA for our theoretical model, which has been described extensively in previous works \cite{maxwell_coulomb-corrected_2017,maxwell_coulomb-free_2018,maxwell_treating_2018,werby_dissecting_2021}, so only a brief overview of the bulk of the method is presented. The technique of unit-cell averaging (first introduced in Ref. \cite{werby_dissecting_2021}) is expanded to allow for in-depth studies of the time separations between interfering trajectories, and so will receive more detailed explanation. 

\begin{figure}
	\centering
    \begin{subfigure}{0.99\linewidth}
    \includegraphics[width=1\linewidth]{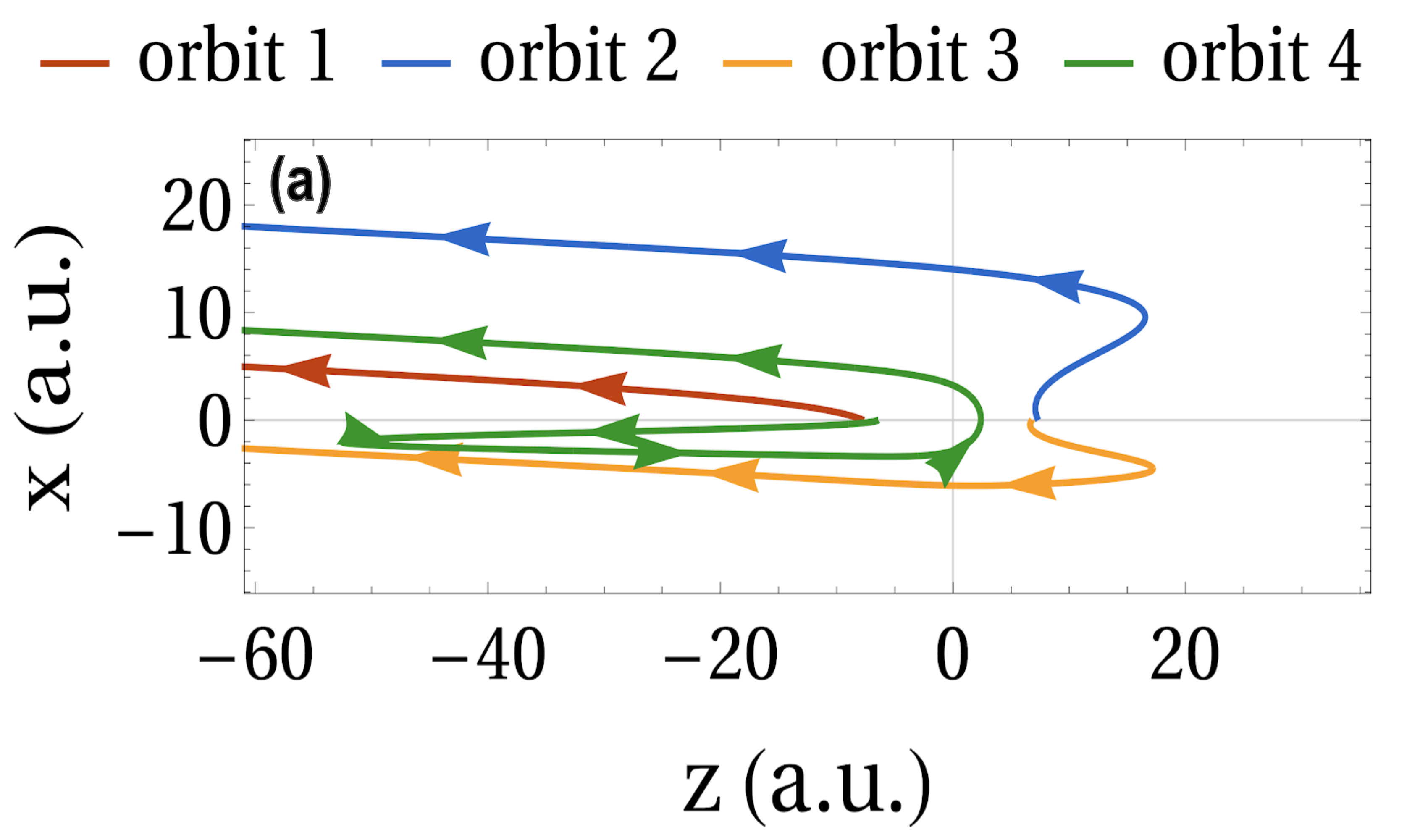}
    \end{subfigure}
    \begin{subfigure}{0.99\linewidth}
    \includegraphics[width=\linewidth]{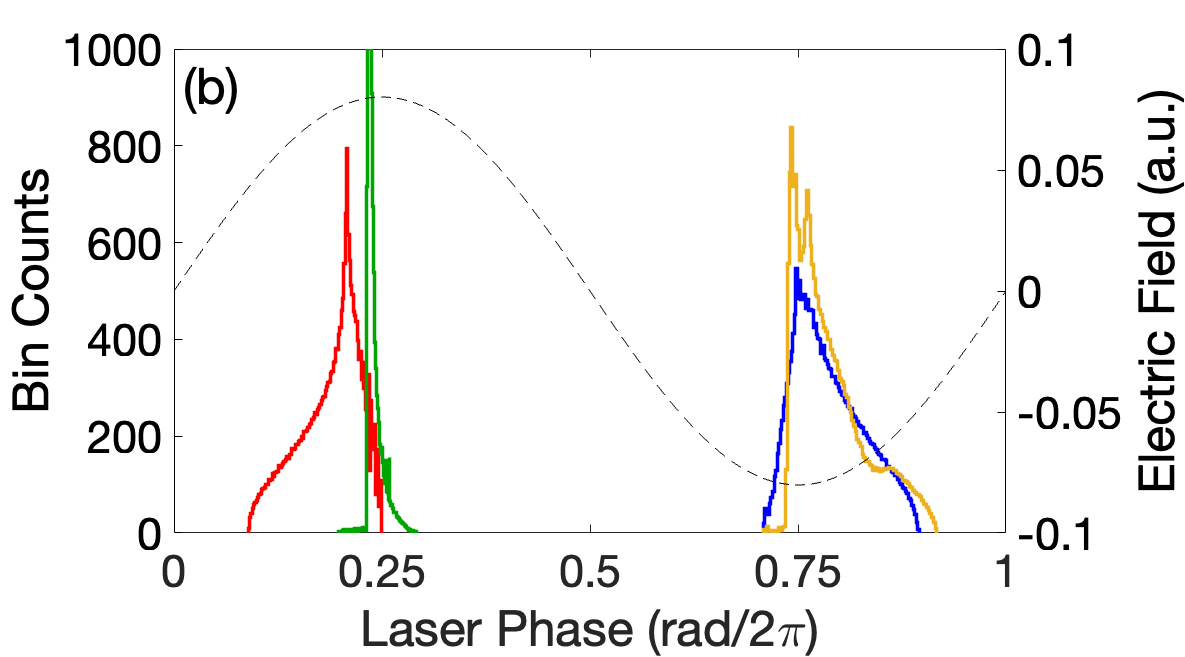}
    \end{subfigure}	
	\caption{(a) Schematic example of the four classifications of orbits identified by CQSFA. The arrows on each trajectory denote the direction of travel, and each arrowhead is separated by 0.2 laser cycles. The position of the parent ion is located at the origin. Each of the shown trajectories has the same final momentum $\pb=(-1.0, 0.13)$, and electrons following them would end up at the same position on the detector. (b) A histogram of the launch times of the orbits identified by the CQSFA. The simulated laser field is overlaid as the black dashed line. These launch times correspond only to the trajectories whose final momenta have a negative parallel component. The launch times for positive momenta are identical with a half cycle shift.}
	\label{fig:CQSFA_Trajectories}
\end{figure}

The starting point for the CQSFA is the transition amplitude $M(\pb)=\braket{\psi_{\pb}|U(t,t_0)}|\Psi_0(t_0)$, where the initial one-electron wave function $\ket{\Psi_0(t_0)}$ is propagated from an initial time $t_0$ to a final time $t$ and projected on to a continuum state $\ket{\psi_{\pb}}$ with a final asymptotic momentum $\pb$. Following the approach in Refs.~\cite{lai_influence_2015,maxwell_coulomb-corrected_2017}, we apply Feynman's path integral formalism \cite{kleinert_path_2009} and the saddle point approximation to the exact transition amplitude given in Ref.~\cite{becker_abovethreshold_2002}. This yields
\begin{align}
&M(\mathbf{p}_f)\propto
-i \lim_{t\rightarrow \infty } \sum_{s}\det \bigg[  \frac{\partial\mathbf{p}_s(t)}{\partial \mathbf{r}_s(t_s)} \bigg]^{-1/2} \hspace*{-0.6cm}
\mathcal{C}(t_s) e^{i
	S(\mathbf{p}_s,\textbf{r}_s,t,t_s))}
\label{eq:MpPathSaddle}
\end{align}
for the transition amplitude, where
\begin{equation}
\label{eq:Prefactor}
\mathcal{C}(t_s)=\sqrt{\frac{2 \pi i}{\partial^{2}	S(\mathbf{p}_s,\textbf{r}_s,t,t_s) / \partial t^{2}_{s}}}\langle \mathbf{p}+\mathbf{A}(t_s)|H_I(t_s)|\Psi_{0}\rangle.
\end{equation}
Here, the interaction Hamiltonian is $\hat{H}_I(t)=\hat{\mathbf{r}}\cdot \mathbf{E}(t)$ and the semi-classical action is given by
\begin{equation}\label{eq:stilde}
    S(\mathbf{p},\textbf{r},t,t')=\ip t'-\int^{t}_{t'}[
    \dot{\mathbf{p}}(\tau)\cdot \mathbf{r}(\tau)
    +H(\mathbf{r}(\tau),\mathbf{p}(\tau),\tau)]d\tau.
\end{equation}
Here, $\ip$ is the ionization potential and the full Hamiltonian
$H(\rb(\tau),\pb(\tau),\tau)=(1/2)\left[\pb(\tau)+\Ab(\tau)\right]^2+V\left(\rb(\tau)\right)$. $V(\rb)$ is the atomic potential of the argon target, for which we use an effective potential provided by Refs.~\cite{kang_holographic_2020,tong_empirical_2005}.

The index $s$ in \eqref{eq:MpPathSaddle} sums over \textit{quantum orbits} that solve the saddle point equations or stationary action. These are given by Newton's equations of motion for continuum propagation and the ionization time equation
\begin{equation}
(\pb(t_s)+\Ab(t_s))^2 + 2\ip = 0.
\label{eq:SPE_times}
\end{equation}
One of the key features for this work is that the laser field, given by $\mathbf{E}(t)=-\partial \Ab(t)/ \partial t$, is taken to be monochromatic, and thus the vector potential is defined to be
\begin{equation}
 \Ab(t) =2\sqrt{\up} \cos(\omega t+\phi),
 \label{eq:AField}
\end{equation}
where $\omega$ is the angular frequency of the laser, $\up$ is the ponderomotive energy, and $\phi$ defines a boundary condition when considering a finite window (or unit cell) from which ionization is allowed to occur. Employing a monochromatic field gives unique control over inter-cycle interferences, which often obscure the intra-cycle holographic interferences of interest.
Here, inter [intra]-cycle interference refers to the combination of two orbits with an ionization times separated by more [less] than a field cycle. \cite{arbo_intracycle_2010}.
In the CQSFA the inter-cycle interference can be ``switched off'' by restricting ionization to a single-laser cycle, which we refer to as a unit cell. The dynamics and phases of the quantum orbits across unit cells will be periodic. However, the parameter $\phi$ controls the ``starting position'' of the laser field in the unit cell.

Solving the saddle point equations lead to four classes of \textit{quantum orbits} \footnote{It is important to note that Maslov phase shifts for specific orbits are accounted for as detailed in Ref.~\cite{brennecke_gouy_2020} and our previous work \cite{werby_dissecting_2021}.}, whose continuum trajectories are plotted in Fig. \ref{fig:CQSFA_Trajectories}(a). 
The trajectories have varying degrees of interaction with the atomic potential: orbit 1 is direct and is only slightly decelerated; orbit 2 is also direct
\footnote{Orbits 1 and 2 are not direct in the same sense of direct SFA orbits, as they interact with the Coulomb potential in the continuum. But both have similar properties to the corresponding direct SFA orbits \cite{maxwell_coulomb-free_2018}}
but is forward deflected as it passes the core; orbit 3 undergoes laser driven forwards rescattering; and orbit 4 is backscattered. Pairwise combinations of these trajectories make many of the well-studied interference patterns in photoelectron holography. This includes the fan-like interference pattern \cite{rudenko_resonant_2004,maharjan_wavelength_2006,gopal_threedimensional_2009,lai_nearthreshold_2017}, with radial fringes that form due to the combination of orbits 1 and 2; the spider-like interference pattern \cite{huismans_time-resolved_2011, hickstein_direct_2012}, with fringes nearly parallel to the polarization axis and form from orbits 2 and 3; and the spiral interference pattern \cite{kang_holographic_2020, maxwell_carpet_2020} that forms from orbits 3 and 4 and has been associated with so-called carpet-like fringes \cite{korneev_interference_2012} that form for large momenta perpendicular to the laser field polarization.

In a previous work \cite{werby_dissecting_2021}, we introduced the technique of unit-cell averaging, which corrected asymmetries generated by selecting an arbitrary value for $\phi$ in a single cycle calculation. Briefly reiterating, when observing the interference between two trajectories within a single cycle, there is an issue with the time ordering of these trajectories. Since each trajectory is periodic, there exists single cycle unit cells such that one trajectory occurs before the other, and other unit cells where it occurs after, having its ionization phase advanced by $2\pi$. When the trajectory advances in this way, it acquires a phase $\Delta S$, given by 

\begin{equation}
    \Delta S = \frac{2\pi}{\omega}\left( 
            \ip + \up +\frac{1}{2}\pb_f^2
    \right),
\end{equation}
where $\frac{1}{2}\pb_f^2$ is the final energy of the electron. Thus, in order to correctly account for the arbitrariness of the unit-cell boundaries, we average the interference over all possible single cycle unit-cells. This is achieved by integrating over $\phi$
\begin{equation}
    I_{\phi}=\int_0^{2\pi} d \phi 
        \exp\left[i \Delta H_{i j}(\phi) \Delta S\right],
        \label{eq:I_Phi}
\end{equation}
where
\[
\Delta H_{i j}(\phi)=H(\phi-\omega t^{\mathrm{Re}}_{i})-H(\phi-\omega t^{\mathrm{Re}}_{j}).
\]
Here $H(t)$ is the Heaviside function, and $t^{\mathrm{Re}}_i$ and $t^{\mathrm{Re}}_j$ are the ionization phases for the trajectories indexed by $i$ and $j$, respectively. The Heaviside functions act to either add or subtract the additional phase $\Delta S$, depending on whether the unit cell beginning at $\phi$ has eclipsed the ionization time of the trajectories indexed by $i$ and $j$, respectively. Then we apply this integral to the sum of the contributions of the orbits:

\begin{equation}
         \mathrm{Prob}(\pb_f)=\frac{1}{2\pi}\sum^4_{i,j=1} M_{i}(\pb_f)\overline{M_{j}(\pb_f)}
        I_{\phi},
        \label{eq:CoherentSum}
\end{equation}
where Prob($\pb_f$) is the probability of an electron arriving at momentum $\pb_f$ which is equivalent to the amplitude at that coordinate in the PMD, $M(\pb_f)$ is the ATI transition element as before, the overline indicates the complex conjugate, and the two sums are each taken over the four orbits identified in CQSFA.






In this paper we introduce a revision to the unit cell averaging which allows us to examine analytically how the time separation affects CQSFA holographic predictions. We are interested in a modification to Eq.~\ref{eq:CoherentSum} such that contributions to Prob($\pb_f$) are coherently summed when the ionization phases of the pair of orbits indexed by $i$ and $j$ are below a given time separation, and incoherently summed when they are above. This must be included within the unit-cell integral in Eq. \ref{eq:I_Phi} and correctly incorporate the unit-cell boundary $\phi$. Thus we arrive at
\begin{align}
I_{\phi}(\sigma)=\int_0^{2\pi} d \phi 
&\exp\left[i \Delta H_{i j}(\phi)\Delta S\right]
\notag\\
        &\times H\left(\sigma-\lvert \omega \Delta t_{ij}+2\pi \Delta H_{i j}(\phi)\rvert\right),
\label{eq:Revised_I_Phi}
\end{align}
where $\Delta t_{ij}=t^{Re}_i-\omega t^{Re}_j$ and we have introduced the parameter $\sigma$ as the threshold time separation between any two orbits. By selecting any specific value for $\sigma$ and then calculating Prob($\pb_f$), we can examine the set of all CQSFA orbits with launch time separations within the window of zero to $\sigma$ cycles apart. We denote this improvement ``windowing'' the CQSFA. Choosing a series of values for $\sigma$ and making a movie of the resulting momentum distributions lets us observe how the interference patterns we observe form. While not further discussed in this paper, a time-separation resolved movie of all the pair-wise CQSFA calculations as well as the full 4-trajectory computation is shown in the supplementary materials for clarity.

\section{Results and Discussion}
\label{Sec:Discussion}

\begin{table*}
    \centering
\begin{tabular}{|M{4cm}|M{3.5cm}|M{2cm}|M{5cm}|}
    \hline
        \textbf{Structure}\vfil &
        \textbf{Trajectory Combination} \vspace{0.8mm}&
        \textbf{Time Separation (rad/$2\pi$)} &
        \textbf{Description}\vfil  \\
        \hline \hline
        ATI Rings \vspace{0.8mm} &
        Any two of the same separated by one cycle &
        1\vspace{0.8mm} &
        Ring-like fringes centered on the origin with photon energy spacing \\
        \hline
        Spider Legs\vspace{0.8mm} &
        2 + 3\vspace{0.8mm} &
        $\sim0$\vspace{0.8mm} &
        Horizontal fringes offset from the polarization axis \\
        \hline
        Carpet\vfil &
        3 + 4\vfil &
        $\sim 0.5$--$0.6$\vfil &
        Checkerboard-like interference pattern near the perpendicular axis \\
        \hline
        Adjacent Rising and Falling Edge Interference Combinations &
        \makecell{1 + 2, 1 + 3, \\ 2 + 4, 3 + 4}\vspace{0.8mm} &
        $0.25$--$0.5$\vfil &
        Pattern of nodes arranged circularly with radial spokes\vspace{0.8mm}\\
        \hline
    \end{tabular}
    \caption{A selection of holographic structures and the trajectories which form them. }
    \label{tab:Holography_Structures}
\end{table*}


In Table \ref{tab:Holography_Structures}, we outline a few predicted holographic structures based on the results of trajectory-based calculations. The pair of trajectories leading to the interference structure are labeled in column 2. In accordance with Fig. \ref{fig:CQSFA_Trajectories}(b), we label the launch time separation between active trajectories in column 3, which is the parameter of interest for us in this paper. Importantly, while these holographic structures have commonly been demonstrated in experiments and the ionization phases of each type of contributing trajectory can be easily calculated, the ionization phases are not in general an accessible observable in these experiments. Thus, these predicted phases have not been empirically verified for these holographic structures. We will demonstrate that the time correlation filter outlined above can reasonably verify the predicted time separations outlined here.

Based on this table, we can set predictions for how the time correlation filter should affect the experimental spectrum if it is successfully filtering on the time separations. First, the null condition: the spider-leg structure. As the spider-leg structure is formed from two electron trajectories passing on either side of the parent ion with nearly overlapping launch times, the time separation between this interfering trajectory pair is close to zero. Thus, the time correlation filter should not affect this structure, and observing Fig. \ref{fig:Exp_Succession}, we see that this is the case for even the most filtered spectrum. The ATI ring structure is also very easy to check. These rings are formed from similar electron trajectories ionized one full cycle apart from each other, and thus the filter should remove them when filtering below one cycle. A previous work was dedicated to selectively removing these rings and uncovering the sub-cycle spectrum beneath them \cite{werby_disentangling_2021}, and it is also clear from the bottom left of Fig. \ref{fig:Exp_Succession}(b) that the ATI rings have been removed for rolloffs below one cycle. 

We now turn our attention to identifying and characterizing the holographic structures corresponding to trajectories launched on opposing half cycles of the laser field (0.5 cycles to 0.75 cycles apart), and on neighboring rising and falling edges (0.25 cycles to 0.5 cycles apart).

\subsection{Confirming the origin of the carpet structure}
\label{sec:Carpet}

\begin{figure}
	\centering
	\begin{subfigure}{0.99\linewidth}
		\includegraphics[width=1\linewidth]{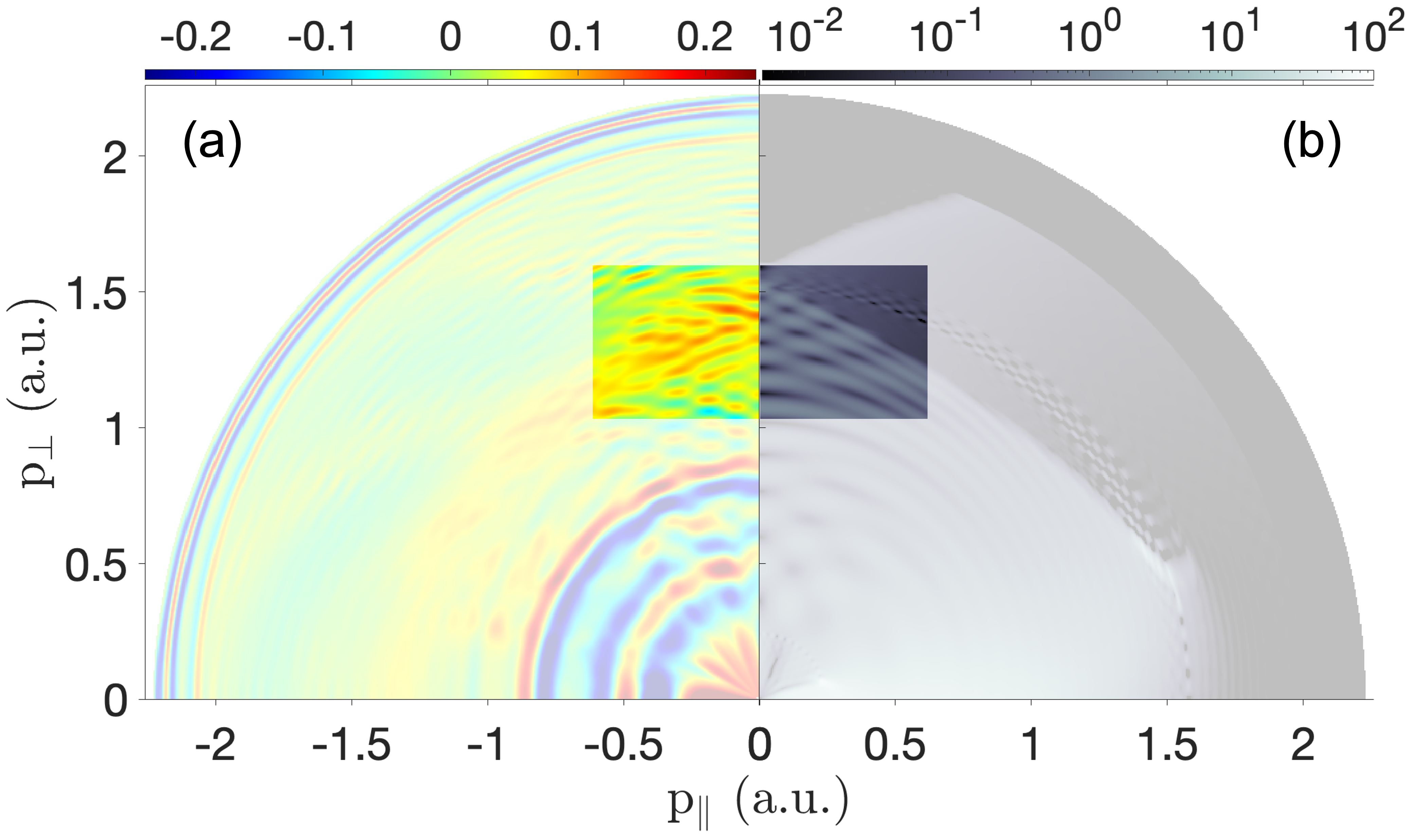}
	\end{subfigure}
	\begin{subfigure}{0.99\linewidth}
	    	\includegraphics[width=1\linewidth]{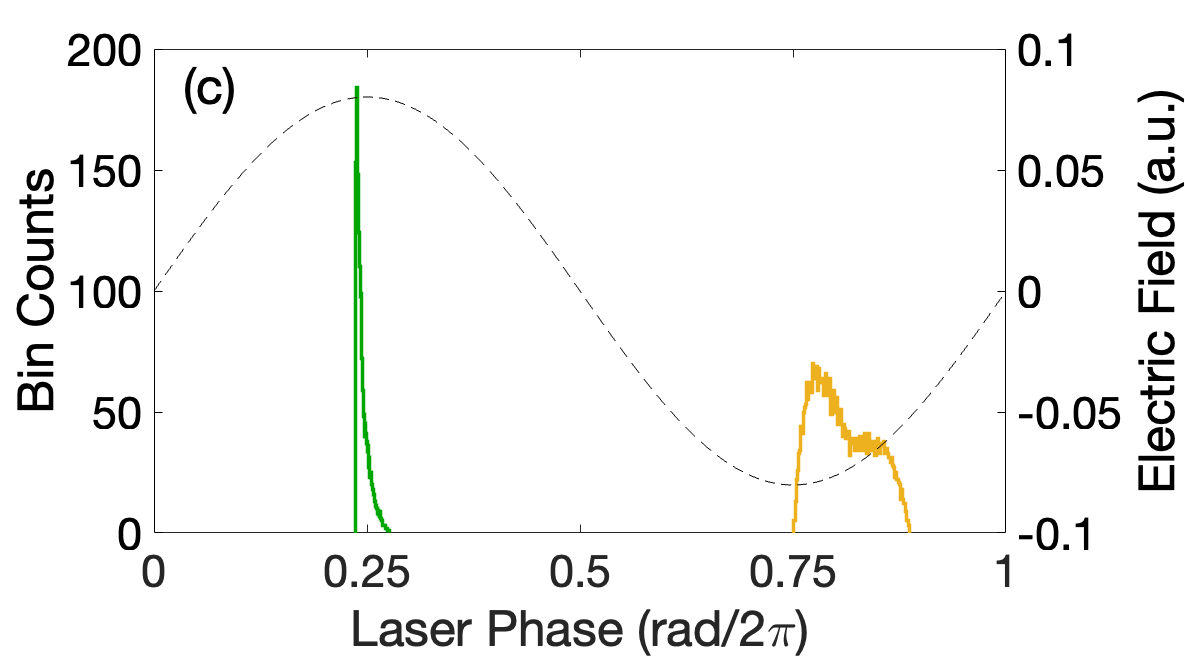}
	\end{subfigure}
	\caption{Comparison of experimentally extracted carpet interference to CQSFA predictions. The box emphasizes the location of the carpet interference. (a) The normalized residual of the experimental spectrum reproduced from Fig. \ref{fig:Exp_Succession}(c) displaying the time separation window of 0.5 to 0.75 cycles. (b) The CQSFA calculation including only the two trajectories forming the carpet structure, 3 and 4. (c) A histogram as in Fig. \ref{fig:CQSFA_Trajectories}(b) except further restricted to just orbits 3 and 4 and the momentum range within the box in the above panel. Once again, only the launch times corresponding to a negative component of parallel momenta are shown.} 
	\label{fig:Carpet_Comparison}
\end{figure}

In the time separation range of 0.5 cycles to 0.75 cycles, one pair of trajectories leads to a very prevalent holographic structure, the carpet. Within the CQSFA framework, the carpet is formed from the interference of electrons following orbits 3 (forward rescattered) and 4 (backward rescattered). In calculation, these trajectories can be isolated and their interference can be computed unambiguously. As the carpet is predicted to form between 0.5 and 0.6 cycles, we expect to observe it in the normalized residuals in Fig. \ref{fig:Exp_Succession}(c). In Fig. \ref{fig:Carpet_Comparison} we show this. In panel \ref{fig:Carpet_Comparison}(a) on the left is the filtered experimental window reproduced from the bottom left of Fig. \ref{fig:Exp_Succession}(c) plotted alongside the CQSFA calculation for the carpet in panel \ref{fig:Carpet_Comparison}(b) on the right. The highlighted box along the vertical axis between the two plots shows the carpet interference. It is clear that the feature revealed in the window exactly matches the predicted shape of this interference. The launch times of the trajectories leading to the carpet structure are presented in panel \ref{fig:Carpet_Comparison}(c), in which only trajectories with final momenta within the white box are considered. From panel \ref{fig:Carpet_Comparison}(c) we see that the carpet structure forms from the interference of orbit 4 launched before orbit 3, such that the time separation is between 0.5 and 0.6 cycles.  

This confirmation of the origin of the carpet structure settles a controversy within the holography community. There has been much debate over the origin of this type of interference, with the trajectory pair depiction being just one of several. Our previous publication \cite{maxwell_carpet_2020} states that the carpet is due to the interference of the CQSFA orbits 3 and 4. This called into question a previous assumption, that the carpet resulted from the interference of the SFA equivalents of orbits 1 and 2 \cite{korneev_carpets_2012}. This previous interpretation reproduces the $2\omega$ spacing in the ATI peaks for a scattering angle perpendicular to the driving-field polarization. However, it does not hold up to scrutiny as it does not match the carpet's energy range and angular behavior observed in experiments \cite{maxwell_carpet_2020}, or the time ranges encountered by the filters in the present work. In fact, a carpet resulting from orbits 1 and 2 would occur in a time separation window of 0.25 to 0.5 cycles apart, and at much lower energy. 


\subsection{Identifying combinations of holograms}

\begin{figure}
	\centering
		\includegraphics[width=1\linewidth]{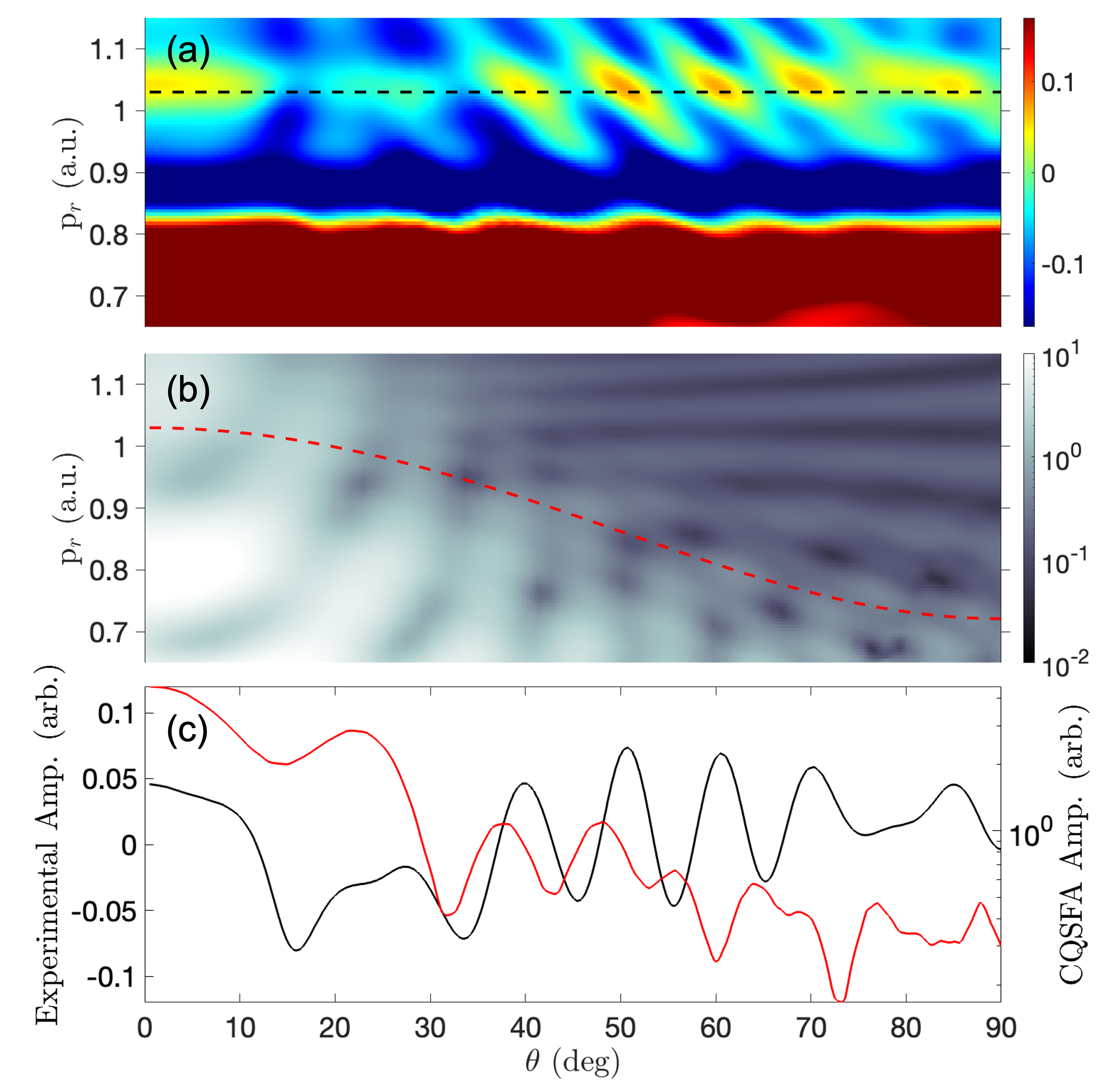}
	\caption{Comparison of the filter window of 0.25-0.5 cycles to the windowed CQSFA calculation of the subset of all interference structures formed in the window of 0.25-0.5 cycles. The outermost direct electrons for experiment and theory are traced in (a) and (b) respectively. Each panel is in polar coordinates, with the shared horizontal axis showing the angle $\theta$ measured from the parallel axis. (a) A section of the experimental window from the bottom left of Fig. \ref{fig:Exp_Succession}(d). (b) The same section of the windowed CQSFA calculation of the structure described in the fourth row of Table \ref{tab:Holography_Structures}. 
	(c) Plot showing the amplitudes of the structures outlined by the black and red dashed lines. The black line uses the y-axis at left and shows the experimental window amplitude, while the red line uses the y-axis at right and shows the windowed CQSFA calculation amplitude.}
	\label{fig:Combination_Comparison}
\end{figure}

The carpet is an excellent two-trajectory feature to identify using the time correlation filter because the primary structure occurs outside the region of the spectrum where direct electrons contribute, and no other holographic structure occurs in that region at the time separation of 0.5 to 0.75 cycles. In general, however, many two-trajectory interference structures do contain direct electrons, and at the time separation of 0.25 to 0.5 cycles, there are many such structures all occurring in the region within 2$\up$. This is because this time separation corresponds to trajectories launched from adjacent rising and falling edges of the field, which predominantly include a direct trajectory and a rescattering one, as can be seen in Fig. \ref{fig:CQSFA_Trajectories}(b). Rather than focus on identifying a single two-trajectory holographic pattern that meets our predictions, we instead must identify a pattern composed of many interference structures forming at this time separation. 

To do this, we examine the predictions of CQSFA to identify the combinations of orbits that will generate their interference structures in this time range. The fourth row of Table \ref{tab:Holography_Structures} lists the combination of CQSFA orbits whose interference leads to structures forming within the range of 0.25 cycles to 0.5 cycles. These trajectories correspond to every combination of orbits omitting the pairs 1 + 4 and 2 + 3 which by observation with Fig. \ref{fig:CQSFA_Trajectories}(b) are nearly overlapping in launch times. Importantly, within this range we are considering the shortest time separation between each of the included trajectory pairs, as every pair has two possible time separations depending on the launch time ordering of the two trajectories. Using the windowed CQSFA, we calculate the PMD using only these contributing trajectories at the short time separation. 

In the experimental filter window between 0.25 cycles and 0.5 cycles (bottom left of Fig. \ref{fig:Exp_Succession}(d)) the most clearly defined structure can be observed along the 2$\up$ boundary. In the experiment this boundary falls along the outermost extent of the direct electrons, which can be seen by observation in Fig. \ref{fig:Exp_Spectrum}. In Fig \ref{fig:Combination_Comparison}(a) the angular distribution of this feature is plotted. The black dashed line traces the centers of the antinodes within the structure and is along the outermost extent of the direct electrons. In Fig. \ref{fig:Combination_Comparison}(b) we present the angular distribution of the CQSFA calculation of the trajectory combination outlined above. Here too the red dashed line traces the outermost direct electrons; however, compared to the experiment there is a known mismatch with the CQSFA because the calculated maximum direct electron energy decreases at higher angle \cite{maxwell_coulomb-corrected_2017}. While there is this discrepancy between the angular behavior of these outermost direct electrons in theory and experiment, a comparison at this boundary is still desired. In \ref{fig:Combination_Comparison}(c) we plot the amplitudes of the spectra along the dashed lines against their angle from the parallel axis. The spacing of the fringes in the compared features closely match, supporting the conclusion that we are observing this predicted combination of interferences in the experimental window.

\subsection{Comparing the time correlation filter to windowed calculations}

\begin{figure*}
	\centering
	\includegraphics[width=1\linewidth]{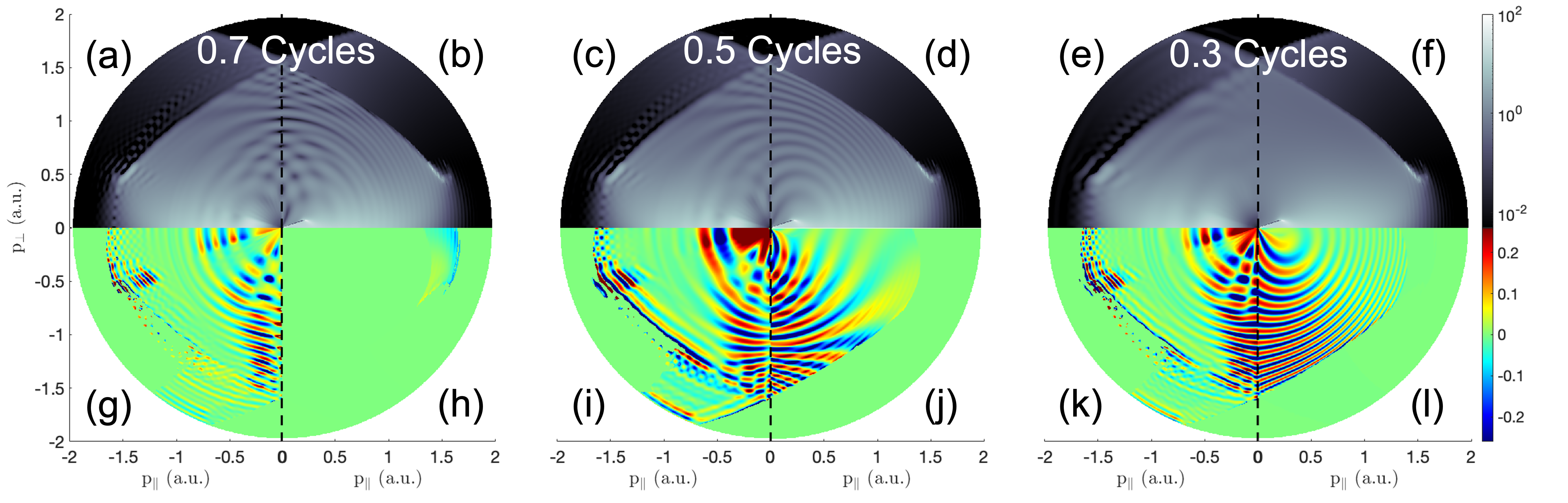}
	\caption{Comparing the time correlation filter on the CQSFA for orbits 3 and 4 (the carpet) to the analytic time separations of windowed unit cell averaging. Here, the time correlation filter has been applied to the CQSFA calculation on the left side of each panel, with the top left showing the filtered calculation and the bottom left showing the filter window between the previous filtered calculation and the one indicated by the annotation at top center. The top right shows the windowed unit cell averaging results, with $\sigma$ equal to the value at top center, and the bottom right shows the normalized residual between the previous and current windowed unit cell averaged results.}
	\label{fig:FilterRUC}
\end{figure*}

At this point it would be prudent to examine the time correlation filter and assess how well it extracts time separations on calculations with known times. For this we compare filtering the CQSFA calculation of only the trajectories forming the carpet to the windowed unit cell averaged calculation at specific time separations. This comparison is shown in Fig. \ref{fig:FilterRUC}. On the left halves of each of the three large plots is the filtered CQSFA calculation of orbits 3 and 4. The top left corners are the filtered CQSFA carpet with a filter rolloff indicated by the annotation at top center, while the bottom left corners are the normalized residuals of the previous less filtered step and the current rolloff as in Fig. \ref{fig:Exp_Succession}. The right halves present the windowed CQSFA orbits 3 and 4 computed by inserting the annotated values for $\sigma$ in Eq. \ref{eq:Revised_I_Phi}. The top right corners are the results of this calculation, whereas the bottom right corners are the normalized residuals with the previous windowed step. The residuals in panels \ref{fig:FilterRUC}(g) and \ref{fig:FilterRUC}(h) are each taken with respect to the unfiltered calculation. 

Comparing the calculations on the right and left in each of the three larger plots we see very good agreement between the filtering method and the windowed calculation. At 0.7 cycles, the carpet interference has fully formed, so we expect the residual from the full calculation to be uniformly zero. Panel \ref{fig:FilterRUC}(g) shows some structure, but overall this window is more suppressed than the other two filtered windows, and the filtered calculation in \ref{fig:FilterRUC}(a) is qualitatively a very good match for the calculation in \ref{fig:FilterRUC}(b). At 0.5 cycles we see that some of the interference has been removed. Consulting Fig. \ref{fig:CQSFA_Trajectories}(b), this interference corresponds to the interference of orbit 4 ionizing before orbit 3. Here we see truly excellent agreement in the shape of the removed interference between panels \ref{fig:FilterRUC}(i) and \ref{fig:FilterRUC}(j), particularly along the vertical axis, indicating that the filter has correctly extracted the same structure that the windowed calculation has. Lastly, at 0.3 cycles of separation, none of the carpet interference has formed, and so the time resolved calculation \ref{fig:FilterRUC}(f) displays no structure. While the filtering has not quite removed all the structure from \ref{fig:FilterRUC}(e), it's clear that the majority of the structure has been suppressed, and once more the residuals \ref{fig:FilterRUC}(k) and \ref{fig:FilterRUC}(l) are in excellent agreement.

The very good agreement between the time correlation filter results and the windowed calculation strongly supports the assertion that the time correlation filter is successfully acting on the previously inaccessible observable of time separation. 


\subsection{Further study}
\label{sec:Further:Study}

In an actual experiment, there are more visible features than can be simply described by approximate theory techniques, no matter how sophisticated. Filtering the raw experimental data reveals concentric interference fringes which move inwards as the filter cutoff is increased. These are not replicated by the windowed CQSFA calculation. However, these 
transient fringes are the strongest feature that we observe in the low momentum region, to the extent where they obscure some of the holographic features we expected to observe within the time window of 0.25 cycles to 0.5 cycles.

\begin{figure}
	\centering
	\includegraphics[width=1\linewidth]{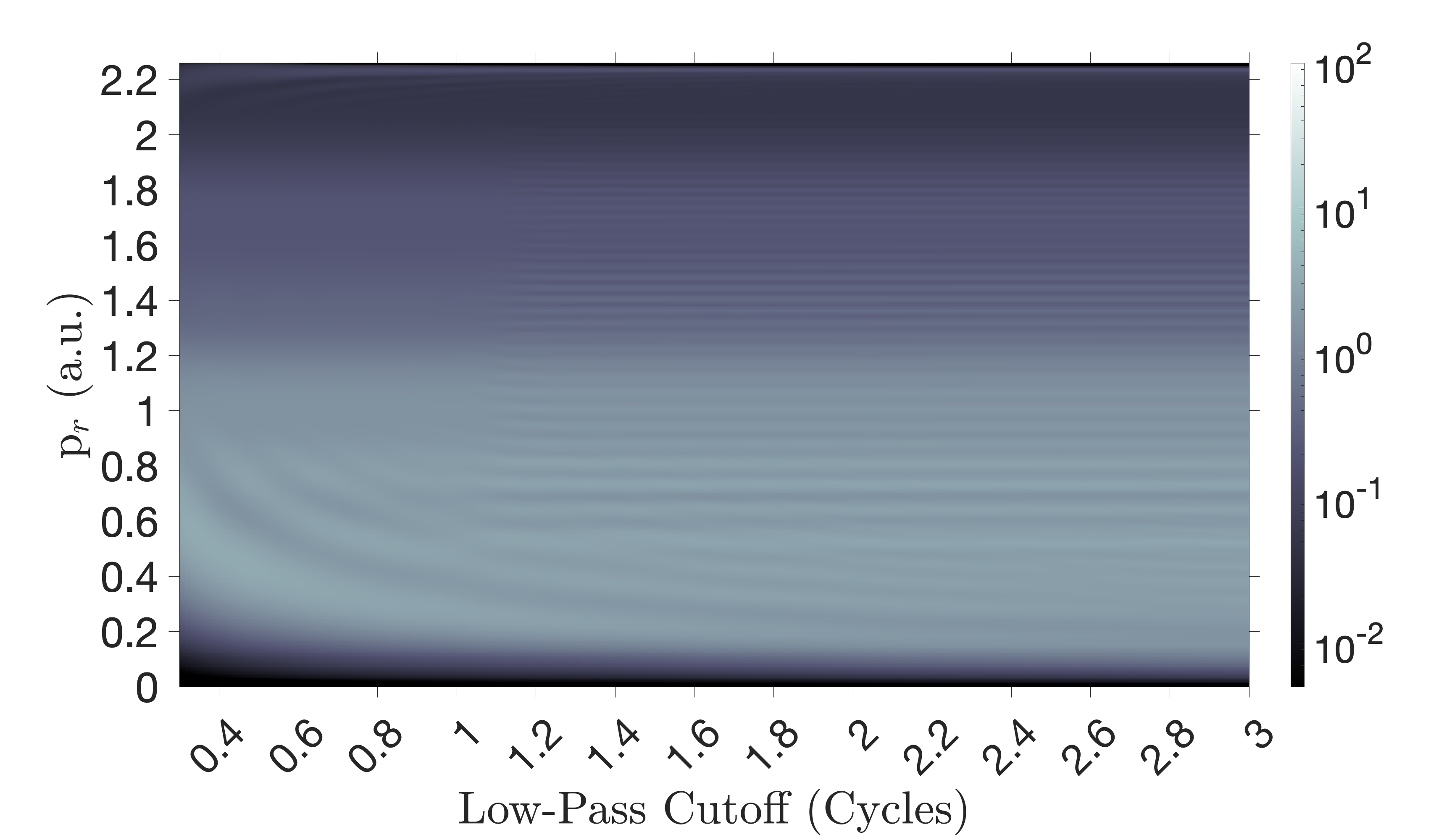}
	\caption{Examining the effect of the transient interference fringes with time correlation filter cutoff. The experimental PMD from Fig. \ref{fig:Exp_Spectrum} has been angularly integrated at each radius $p_r$ to capture the behavior of rings about the origin, and then filtered at fine cutoff steps up to a maximal cutoff of 3 cycles.}
	\label{fig:TransientRings}
\end{figure}

In Fig. \ref{fig:TransientRings} we examine the behavior of these transient fringes by plotting the integrated electron yield along successive rings of radius $p_r$ against the filter cutoff. This is a useful representation to demonstrate the ATI ring structure, which forms horizontal lines beginning at a filter cutoff of one cycle. The transient fringes are the curved features below $p_r = 1$ a.u. It is not immediately apparent what the source of these fringes might be; however, this pattern is very dominant and highly structured within this region of the spectrum, and thus likely has some simple explanation. 

It is important to note here that the filtering technique is essentially just a different way of viewing the same data to uncover features already present in the spectrum. The emergence of transient patterns that elude explanation when viewing the data in this way is an intriguing conundrum and merits further study. 

\subsection{Limitations of the time correlation filter}
\label{sec:Limitations}

Many of the above determinations implicitly relied on the fact that holographic interference features have fringe spacings which are dominated by a term proportional to $E(t_1-t_2)$, i.e. linear in the ionization time difference. When this is the dominant term, the time correlation filter is very accurate, and is able to reproduce holographic features within a narrow time window. This is expected for any pair of interfering trajectories around the perpendicular axis ($\theta\approx\pi/2$) and is particularly clear with the carpet structure, which displays this behavior in this region. 
Away from this region, the CQSFA (and even the SFA) predict the inclusion of interference terms which are not linear in the ionization time difference. This means the temporal information yielded by the cosine transform does not exactly correspond to the ionization time difference predicted by the CQSFA. 
Calculating this mismatch in the worst case at $\theta=0$ yields a maximum error of $0.25$ cycles, which is within the bounds of error required for the conclusions drawn in this manuscript.

\section{Conclusion}

We have developed a powerful analysis technique for observing electron trajectory interference in SFI. By analyzing the cosine transforms of the anisotropy parameters of a VMI spectrum, we identify trajectory pair time correlations. Filtering these quantities at successive rolloffs uncovers trajectory pair interferences and provides an experimental determination of the launch time separations of some of the more frequently discussed holographic features. We compare directly to CQSFA calculations of the full spectrum, as well as to specific combinations of electron trajectories and see dramatic agreement. In particular, the carpet interference structure which is normally difficult to observe in ordinary VMI spectra can both be extracted clearly by the time correlation filter in the predicted range of 0.5 to 0.75 cycles and matched exactly to the CQSFA prediction of the electron trajectories forming the carpet. This experimentally confirms the prediction that the carpet interference forms from the interference of forward and backward rescattered trajectory pairs, which up until only recently had been debated within the holography community.

In order to test that the time correlation filter effectively extracts the launch time separations of interfering electron trajectory pairs, we developed an advancement to the CQSFA technique of unit-cell averaging to calculate the spectrum for selected time separations. Comparing filtered CQSFA calculations to these windowed spectra shows excellent agreement, and confirm that the correct interference structures are extracted at the proper times. 

The power of the technique lies in its applicability. As an analysis technique, time correlation filtering can be applied to all energy- and angularly-resolved spectra, and can be applied retroactively to analyze previously collected data. Using the filter to uncover time correlations complements other experimental designs seeking to explore ionization times in strong fields, such as phase of the phase. This filter is a powerful and versatile tool which can provide additional insight for many types of strong-field experiments, without requiring any changes in experimental design or technique.

\begin{acknowledgments}

NW, RF, and PHB are supported by the U.S. Department of Energy, Office of Science, Basic Energy Sciences (BES), Chemical Sciences, Geosciences, and Biosciences Division, AMOS Program. RF gratefully acknowledges support from the Linac Coherent Light Source, SLAC National Accelerator Laboratory, which is supported by the US Department of Energy, Office of Science, Office of Basic Energy Sciences, under contract no. DE-AC02- 76SF00515.
ASM acknowledges funding support from the European Union’s Horizon 2020 research and innovation programme under the Marie Sk\l odowska-Curie grant agreement, SSFI No.\ 887153 and when part of the QOT group in ICFO acknowledges support from
Agencia Estatal de Investigaciín (the R\&D project CEX2019-000910-S, funded by MCIN/ AEI/10.13039/501100011033, Plan National FIDEUA PID2019-106901GB-I00, FPI), Fundació Privada Cellex, Fundació Mir-Puig, and from Generalitat de Catalunya (AGAUR Grant No. 2017 SGR 1341, CERCA program). CFMF was funded by grant No.\ EP/J019143/1, from the UK Engineering and Physical Sciences Research Council (EPSRC).

\end{acknowledgments}


\bibliography{main}{}
\bibliographystyle{apsrev4-2}

\end{document}